# Structural, dielectric, electrocaloric, and energy storage properties of lead free $Ba_{0.975}La_{0.017}(Zr_xTi_{0.95-x})Sn_{0.05}O_3$ (x = 0.05; 0.20) ceramics

Sihem Smail[a], Manal Benyoussef[b], Kamel Taïbi[a], Bouchaib Manoun[c,d], Mimoun El Marssi[b], Abdelilah Lahmar[b,*]

[a]*Laboratoire de Cristallographie-Thermodynamique, Faculté de Chimie, U.S.T.H.B., BP32, Al Alia, 16111, Alger, Algeria.*
[b]*Laboratoire de Physique de la Matière Condensée (LPMC), Université de Picardie Jules Verne, 33 rue Saint-Leu, 80039 Amiens Cedex 1, France*
[c]*Laboratoire de Rayonnement-Matière et Instrumentation, S3M, FST Settat, Université Hassan 1er, Morocco.*
[d]*Materials Science and Nano-engineering, Mohammed VI Polytechnic University, Lot 660 Hay Moulay Rachid, Ben Guerir, Morocco.*

**Abstract**

A-site deficient $Ba_{0.975}La_{0.017}\square_{0.0085}(Zr_xTi_{0.95-x})Sn_{0.05}O_3$ (x = 0.05; 0.20) ceramics (BLaZ100xTS) were synthesized using the solid-state reaction. The microstructural study revealed a high density and low porosity in the studied ceramics. Besides, grain size lower than 1.5 µm was observed in the studied compounds, which is due to the incorporation of the rare earth element (lanthanum). X-ray diffraction and Raman studies revealed that BLaZ5TS and BLaZ20TS crystallize in a perovskite-type structure with tetragonal and cubic symmetry, respectively. The dielectric study showed a normal ferroelectric-paraelectric phase transition for BLaZ5TS while a diffuse phase transition is noticed for BLaZ20TS sample.

The energy-storage density and the associated energy efficiency were determined from the P-E loops versus temperature and the calculated values are comparable with results obtained on BCZT systems. Furthermore, the adiabatic temperature change ΔT was calculated through the indirect method and a relatively high value of $\Delta T/\Delta E = 0.2 \cdot 10^{-6}$ K m/V is obtained in BLaZ5TS system. The simultaneous presence of energy storage property and electrocaloric responsivity makes this system a promising environmentally friendly candidate for application in electronic devices.

Keywords: BZT-based ceramics; dielectric investigations; energy storage; electrocaloric effect

*Corresponding author: abdel.ilah.lahmar@u-picardie.fr

# 1. Introduction

In recent decades, components involved in electronic equipment became increasingly sophisticated to meet the requirements of emerging technologies. The literature survey showed that several materials have been explored [1–4]. Besides, ferroelectric perovskite materials are the most promising candidates for low-cost electronic components due to their easy manufacturing as well as their rich functional properties [5–10].

Unfortunately, perovskite materials used actually in the manufacture of the electronic devices are mainly lead-based compounds such as $PbMg_{1/3}Nb_{2/3}O_3$ (PMN), $PbSc_{1/2}Nb_{1/2}O_3$ (PSN) and $Pb(Zr_xTi_{1-x})O_3$ (PZT). Nevertheless, the toxicity of lead introduces inherent health risks and environmental damage. In this context, a universal industrial restriction has recently been adopted regarding the use of toxic materials [11]. Consequently, the development of both efficient and environmentally friendly materials has become an alternative arousing great attention. For instance, $Ba(Zr_xTi_{1-x})O_3$ (BZT) systems are recognized as eco-friendly lead-free ferroelectrics with potential applications involving piezoelectric actuators, multilayer ceramic capacitors, tunable microwave devices, etc. [12–17]. Generally, the performances of these materials are mainly related to the significant changes in the electrical and structural properties, generated by the partial replacement of $Ti^{4+}$ by $Zr^{4+}$ in BZT matrix.

In the last decade, new emerging applications in BZT-based compounds such as electrocaloric cooling systems or energy storage capacitors have seen a number of publication soaring [18–26]. Asbani et al. [18] reported a large electrocaloric responsivity coefficient ($\Delta T/\Delta E$) of 0.34 K mm/kV in $Ba_{0.8}Ca_{0.2}Zr_{0.04}Ti_{0.96}O_3$ peaked at 386K. However, Kaddoussi et al. [19] showed that the presence of $Sn^{4+}$ depressed the transition temperature without affecting the value of the ECE responsivity. In a subsequent work, the authors showed that the simultaneous presence of Ca and Sn in the $Ba(Zr_{0.1}Ti_{0.9})O_3$ matrix-induced a diffuse ferroelectric phase transition with an enhancement of the electrocaloric response in a broad temperature range surrounding room temperature [20]. In fact, the adjustment of the transition temperature is possible using the incorporation of $Sn^{4+}$ ions in the BZT matrix [19,20].

On the other hand, the energy storage properties in BZT-based ceramics have received also great attention in both bulk [27] and thin-film forms [24–26]. It has been shown that these systems have the potential ability to deliver high power density with an ultra-fast charging-discharging process, promising for advanced electronic applications such as in electric power systems, or in high-frequency inverters and power grids [28,29].

Recall that only antiferroelectric (AF) and relaxor ferroelectric (RF) materials endorsing low remanent polarization, high saturation polarization in addition to the low dielectric loss, are judged able to present good energy storage properties. However, most of the AF are Pb-based material. In the case of BZT systems, the variation of Zr composition, as well as the introduction of Sn into Ti- site promotes the diffuse phase transition and therefore the relaxor behavior [30–32].

In the present work, we relate and discuss the structural, dielectric, energy storage property, and electrocaloric effect of $Ba_{0.975}La_{0.017}\square_{0.085}(Zr_xTi_{0.95-x})Sn_{0.05}O_3$ (symbolized BLaZ100xTS) where x represents low (0.05) and high (0.20) Zr contents. The low Zr content (x = 0.05), and high Zr content (x = 0.20) were chosen in order to investigate systems in both normal ferroelectric, and relaxor ferroelectric order, respectively. It should be noted that the La and Sn contents for this study were fixed at 0.025 and 0.050, respectively.

The selection of this trivalent lanthanide is motivated by its particular electronic configuration likely to induce specific behaviors. Indeed, the trivalent $La^{3+}$ ion behaves as a donor by replacing $Ba^{2+}$ ions in the BZT compounds, leading to a charge imbalance, which promotes the ferroelectric relaxor effect suitable for ECE and energy storage improvement.

## 2. Materials and methods

2.1 Solid-state synthesis

$Ba_{0.975}La_{0.017}\square_{0.085}(Zr_xTi_{1-x})_{0.95}Sn_{0.05}O_3$ (x = 0.05; 0.20) compositions were synthesized using conventional mixed oxide method according to the following reactions:

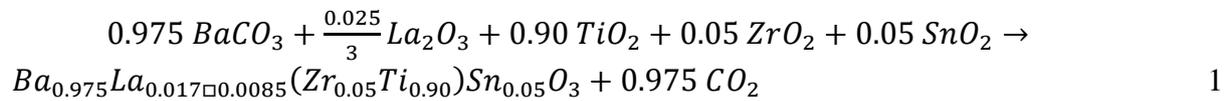

$$0.975\ BaCO_3 + \frac{0.025}{3} La_2O_3 + 0.90\ TiO_2 + 0.05\ ZrO_2 + 0.05\ SnO_2 \rightarrow$$
$$Ba_{0.975}La_{0.017}\square_{0.0085}(Zr_{0.05}Ti_{0.90})Sn_{0.05}O_3 + 0.975\ CO_2 \qquad 1$$

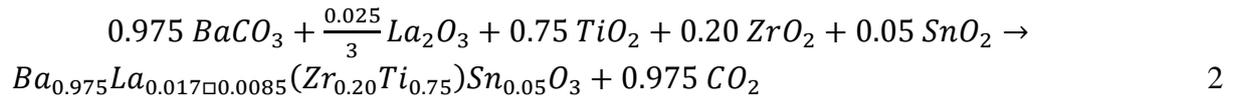

$$0.975\ BaCO_3 + \frac{0.025}{3} La_2O_3 + 0.75\ TiO_2 + 0.20\ ZrO_2 + 0.05\ SnO_2 \rightarrow$$
$$Ba_{0.975}La_{0.017}\square_{0.0085}(Zr_{0.20}Ti_{0.75})Sn_{0.05}O_3 + 0.975\ CO_2 \qquad 2$$

where "$\square$" represents the vacancies created during the reaction at high temperature.

High purity raw materials (>99.99%) were used for the synthesis. Stoichiometric amounts of reagents $BaCO_3$, $La_2O_3$, $TiO_2$, $ZrO_2$, and $SnO_2$ were weighed in molar proportions, mixed and milled for 2 h in an agate mortar. Afterward, they were heat-treated under an air atmosphere at 1170 °C for 24 h. After new intimate and ground mixings, the mixture was then pressed under 500 MPa into a pellet of 13 mm of diameter and about 1 mm of thickness. The resulting disk-shaped ceramics were sintered in air for 4 h at 1400 °C. Loss of weight was determined before and after heat treatment and was found less than 1%. Diameter shrinkages $\Delta\Phi/\Phi$ were determined as $(\Phi_{initial} - \Phi_{final})/\Phi_{initial}$. Their values were about 10% while the relative density (experimental density/theoretical density) was about 0.90%.

2.2. Characterizations

The identification of the formed phase, the related symmetry, as well as the unit-cell parameters were achieved using a D8 Advance X-ray diffractometer (Vantack detector). The data were collected at room temperature using CuK$\alpha$1+2 radiation ($\lambda$= 1.541 Å) in the 2θ range from 20° to 80°. The phase identification was done by comparison of the diffraction patterns with the reference cards of the JCPDS Powder Diffraction File. The XRD analysis was achieved by the Rietveld refinement of the ceramics structure using the Fullprof program. A pseudo-Voigt profile function was used to describe the peak shape, starting with atomic positions taken from $BaTiO_3$ [33].

The microstructures of the sintered samples were examined by scanning electron microscopy analysis using a Philips SEM device (model XL30).

Dielectric measurements were performed on the ceramic discs after DC sputtering of platinum electrodes on the circular faces. The dielectric permittivity and loss were measured as a function of both temperature (100-450 K) and frequencies ($10^2$-$10^6$ Hz) using a Solartron Impedance Analyser SI 1200. The ferroelectric hysteresis loops were determined at different temperatures using a ferroelectric test system (TF Analyzer 3000, aixACCT) and a custom-made heating stage. The Raman spectra were recorded using a micro-Raman Renishaw spectrometer equipped with a CCD detector and a Linkam TS-93 stage allowing temperature stability of ± 0.1 K. With this device, it is possible to conduct measurement at low and high temperature going from 400 to 570 K. Raman spectra were recorded at room temperature using a micro-Raman Renishaw spectrometer equipped with a CCD detector. The green laser was used to excite the samples (514.5 nm).

**3. Results and discussion**

3.1. Microstructure investigation

Figure 1 shows the SEM micrographs of both BLaZ5TS and BLaZ20TS samples. As observed, the grains have spherical shapes with a relatively small grain size. Besides, high microstructural density and low porosity are evidenced in the samples. It seems that $La^{3+}$ ions are mainly incorporated in the A site, act as a donor, and slow down the diffusion during the sintering process. Therefore, a low grain size below 1.5 µm is obtained (see inset in figure 1 (a,b)). Furthermore, compared to the microstructure of BZT compounds reported in refs. [34,35], grain growth seems to be inhibited by the addition of lanthanum. Indeed, the average grain size in undoped BZT (> 10 µm) is much larger than that observed in BLaZ5TS and BLaZ20TS samples (0.2 – 1.5 µm).

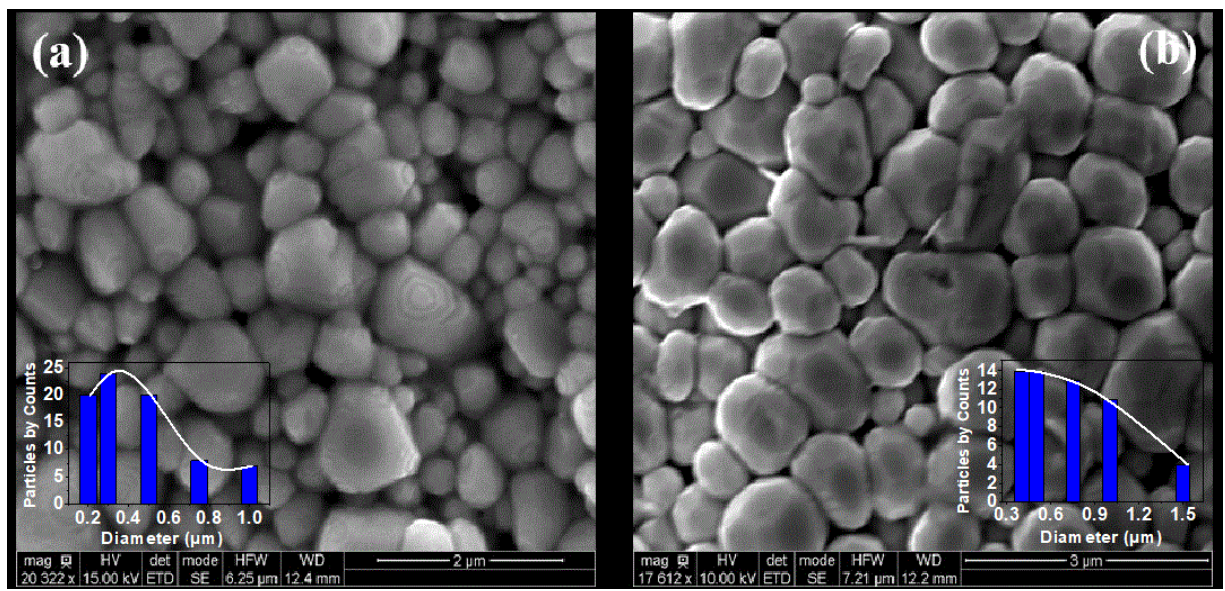

**Figure 1:** SEM images for (a) BLaZ5TS, (b) BLaZ20TS. The inset shows the histogram of grain size distribution.

3.2. Room-temperature X-ray diffraction and Raman analysis

The phase formation of the synthesized BLaZ5TS and BLaZ20TS ceramics was confirmed by the X-ray diffraction technique at room temperature. The examined compositions are marked by the change in the Zr content from x = 0.05 (BLaZ5TS) to x = 0.20 (BLaZ20TS) and correspond to the substitution of $Ti^{4+}$ by $Zr^{4+}$. The patterns were refined using the FullProf program [36] integrated into the WinPLOTR software [37]. The Rietveld refinement results and refinement reliability factors were achieved assuming a tetragonal symmetry (*P4mm*, JCPDS No. 83-1860) for BLaZ5TS and a cubic symmetry (Pm-3 m, JCPDS No. 75-0461) for BLaZ20TS. Fig.2 (a-b) illustrates a typical Rietveld refinement pattern at room temperature for both samples. Note that the XRD patterns contain $La_2Zr_2O_7$ impurity. The amount of the impurity was about 2.1% for BLaZ5TS phase leading to the crystallographic formula: $Ba_{0.9755}La_{0.01633}Zr_{0.049}Ti_{0.901}Sn_{0.05}O_3$. For BLaZ20TS composition, the amount of the impurity was about 1.5% leading to the crystallographic formula $Ba_{0.9754}La_{0.0164}Zr_{0.197}Ti_{0.753}Sn_{0.05}O_3$.

Table 1 gathered the obtained symmetries and lattice parameters comparatively to $BaTiO_3$ as a reference. It was found that the composition with higher Zr content has greater lattice parameters (a) and volume (V). These variations are undoubtedly related to the bigger size of $Zr^{4+}$ than that of $Ti^{4+}$ ($r_{Ti4+}$ = 0.605 Å and $r_{Zr4+}$ = 0.720 Å in 6 Coordination Number [38]). Consequently, the volume and the crystal lattice parameters increased while substituting $Ti^{4+}$ by $Zr^{4+}$. In addition, the values of the c/a ratio observed in tetragonal symmetry are very close to unity involving pseudo-cubic symmetry for BLaZ5TS phase. This assertion is in good agreement with the dielectric study that will be presented in Sect. 3.3.

The atomic positions, occupancy and isotropic atomic displacement parameters of the perovskite compositions BLaZ5TS and BLaZ20TS as well as the Rietveld refinement reliability factors are summarized in Tables 2 and 3, respectively. From the analysis of the various inter-atomic distances, it was observed that $Ti^{4+}/Zr^{4+}/Sn^{4+}$ cations at the B-sites are octahedrally coordinated with the oxygen atoms to form $(Ti/Zr^{4+}/Sn^{4+})O_6$ octahedra. These octahedra are connected by oxygen atoms to form a three-dimensional network. We found that $(Ti/Zr^{4+}/Sn^{4+})O_6$ octahedra are only distorted in the *P4mm* space group.

On the other hand, it should be noted that the $R_{wp}$ values are relatively higher compared to the usual reliability factors. However,, these values are reasonable for disordered perovskites, such is the case for $PbMg_{1/3}Nb_{2/3}O_3$ compound (PMN) well-known with the nanoscale atomic disorder [39]. Noting that such type of disorder is reported to be at the origin of the observation of the cubic symmetry in both ferroelectric and paraelectric phases in the case of relaxor ferroelectric materials [1,40].

**Table 1:** Symmetry and lattice parameters of BLaZ5TS, BLaZ20TS, and $BaTiO_3$ ceramics.

|  | Lattice parameters | | | |
| --- | --- | --- | --- | --- |
| Composition and symmetry | a(Å) | c(Å) | c/a | Volume(Å$^3$) |
| BLaZ5TS (Tetragonal) | 4.0517 | 4.0456 | 0.9985 | 66.41 |
| $BaTiO_3$ (Tetragonal)[a] | 3.9945 | 4.0335 | 1.0098 | 64.36 |
| BLaZ20TS (Cubic) | 4.0807 | - | - | 67.95 |
| $BaTiO_3$ (Cubic)[b] | 4.0119 | - | - | 64.57 |

[a] PDF card 83-1860, JCPDS     [b] PDF card 75-0461, JCPDS

**Table 2:** Rietveld refinement results

| Composition | Ions | Site | x | y | Z | $B_{iso}$ | Occupancy |
| --- | --- | --- | --- | --- | --- | --- | --- |
| BLaZ5TS | $Ba^{2+}/La^{3+}$ | 1a | 0 | 0 | 0.0004 (5) | 0.45 (2) | 0.9755/0.01633 |
|  | $Ti^{4+}/Zr^{4+}/Sn^{4+}$ | 1b | 0.5 | 0.5 | 0.5079 (3) | 0.83 (3) | 0.901/0.049/0.05 |
|  | $O_1^{2-}$ | 1b | 0.5 | 0.5 | -0.0368 (5) | 1.02 (5) | 1 |
|  | $O_2^{2-}$ | 2c | 0.5 | 0 | 0.4796 (4) | 1.02 (5) | 2 |
| BLaZ20TS | $Ba^{2+}/La^{3+}$ | 1a | 0 | 0 | 0 | 1.114 (3) | 0.9753/0.01644 |
|  | $Ti^{4+}/Zr^{4+}/Sn^{4+}$ | 1b | 0.5 | 0.5 | 0.50 | 1.284 (2) | 0.752/0.198/0.05 |
|  | $O_1^{2-}$ | 3c | 0 | 0.5 | 0.5 | 1.003 (4) | 3 |

**Table 3:** Refinement reliability factors

| Composition | $R_B$ (%) | $R_F$ (%) | $R_{wp}$ (%) | $\chi^2$ |
|---|---|---|---|---|
| BLaZ5TS | 3.63 | 2.56 | 14.93 | 2.79 |
| BLaZ20TS | 4.03 | 3.51 | 15.56 | 3.22 |

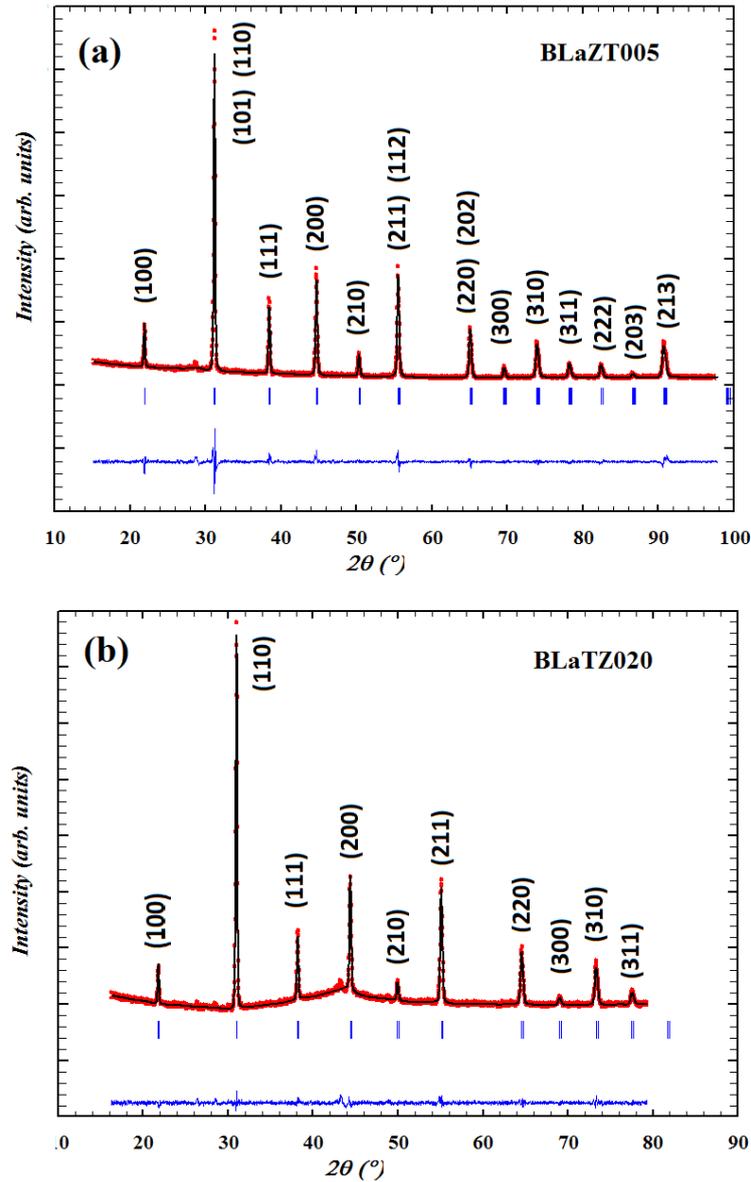

**Figure 2:** Final Rietveld refinement plots of X-ray diffraction patterns for (a) BLaZ5TS and (b) BLaZ20TS. Red open circles indicate the observed data and black solid lines indicate the calculated pattern with Rietveld method. The lower curve is the difference between observed and calculated pattern. The calculated positions of Bragg reflections are marked by vertical markers.

Figure 3 shows the Raman spectra recorded at room temperature for the BLaZ5TS and BLaZ20TS, compounds. These spectra are made up of six bands, which can be divided into 2 categories, namely LO (Longitudinal Optic) and TO (Transverse Optic) [41,42]. The spectra of low Zr content (BLaZ5TS) are distinguished by modes characteristic of the tetragonal structure. Noting that these modes, appearing at 188, 261, 306, 515, and 720 cm$^{-1}$, are different from those related to the compositions with higher Zr content (BLaZ20TS) as we can see in the region 180-308 cm$^{-1}$. This result confirms the structural transition from tetragonal to cubic symmetry highlighted by X-ray diffraction analysis and associated with the change in Zr concentrations from 0.05 to 0.20. On the other hand, the modes $A_1(TO_1)$ and $A_1(TO_2)$ observed respectively at 188 and 261cm$^{-1}$ are attributed to the vibration of symmetrical stretching and relative to the deformed octahedral $BO_6$ clusters (B = Ti, Zr, Sn) [43,44]. In addition, the $A_1(TO_3)$ mode at 515 cm$^{-1}$ is associated with asymmetric vibrations, and the $B_1$ mode located at 306 cm$^{-1}$ with an overlap of E(TO) and E(LO) indicates an asymmetry in the $BO_6$ octahedra [45]. The $A_1(LO_3)$ mode observed at 720 cm$^{-1}$ is analogous to the mode appearing when zirconium (Zr) is substituted by titanium (Ti). This mode has been attributed to the tetragonal symmetry of the BZT and could be due to the propagation of phonons along the c axis [46]. Concerning the Raman spectra of the higher Zr composition (BLaZ20TS), the appearance of a sharp peak around 112 cm$^{-1}$ associated with the mode $E(TO_1)$ is noticed, and corresponds to the transition from tetragonal to cubic phase [47]. Besides, we note that the intensity of the $A_1(LO_3)$ mode decreases significantly leading to a broad peak. This is in good agreement with previous works claiming the absent of such feature in the paraelectric cubic phase [48,49].

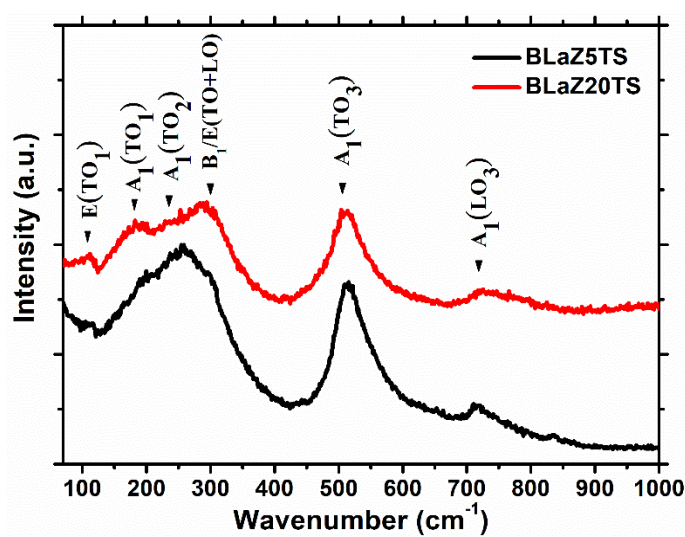

**Figure 3:** Raman spectra recorded at room temperature for the BLaZ5TS, BLaZ20TS compounds.

3.3. Dielectric study

Figure 4 (a,b) presents the temperature and frequency dependence of the dielectric permittivity and losses at different frequencies for BLaZ5TS and BLaZ20TS samples. Two different behaviors were evidenced. For the composition with low Zr content (BLaZ5TS, Figure 4 (a)), three phase transitions can be observed, similar to the normal ferroelectric $BaTiO_3$ (BT) system. Recall that the three phase transitions in BT from rhombohedral-to-orthorhombic (R-O), orthorhombic-to-tetragonal (O-T), and tetragonal-to-cubic (T-C) take place at temperatures of $T_{R-O}$ = 183 K, $T_{O-T}$ = 268 K, and $T_{T-C}$ = 393 K, respectively [50]. For BLnZ5TS ceramic (figure 4) the T-C phase transition is shifted to low temperature around 320 K while the ferroelectric R-O and O-T phase transitions are shifted to high temperature at 250 K and 292 K, respectively. The resulted gathering of the phase transitions around room temperature in this composition is highly desired for high electrocaloric response in a broad temperature range.

However, unlike BT, here the $T_{R-O}$ and $T_{O-T}$ manifested by a slight bump in the dielectric permittivity but can be clearly observed in the dielectric losses as shown in the inset of figure 4 (a). Besides, the sharp peak observed usually at $T_{T-c}$ in pure BT, turned into a relatively broad peak in the case of BLaZ5TS compound. A similar result has been reported by other authors, claiming that the incorporation of lanthanide elements in the BZT compositions with low Zr content, exhibited a normal ferroelectric behavior with a diffuse character [51–53].

Regarding the composition with relatively high Zr content (BLaZ20TS, Figure 4(b)), a diffuse phase transition without frequency dispersion is observed. Consequently, a very broad peak took place where the maximum permittivity ($T_m$) appeared to be frequency independent.

Furthermore, it should be noted that low dielectric losses have been highlighted for both BLaZ5TS and BLaZ20TS compositions. The obtained low losses are principally due to the resulted dense microstructure in our samples. Besides, rare earth doping into ferroelectric perovskites is generally reported to result in improved dielectric losses [54,55]. For instance, $Gd^{3+}$ substitution into the A-site of $BaZr_{0.05}Ti_{0.95}O_3$ system was found to result in reduced dielectric losses, which was mainly attributed to the effect of rare earth on the reduction of titanium vacancies [51]. Interestingly, the investigated compositions in the present work present considerably low dielectric losses compared to the pure BZT system. Besides, the addition of impurities into ferroelectric perovskites was revealed to reduce considerably the dielectric losses, which is beneficial for use in tunable radio frequency and microwave devices [56]. We believe that the presence of $La_2Zr_2O_7$ as a secondary phase plays a key role in the reduction of the dielectric losses in our samples.

In particularly, at low frequency and between room temperature and 450°C extremely low dielectric losses below 0.02 (i.e. 0.01 at RT) are obtained for the BLaZ5TS composition, which is suitable for practical applications.

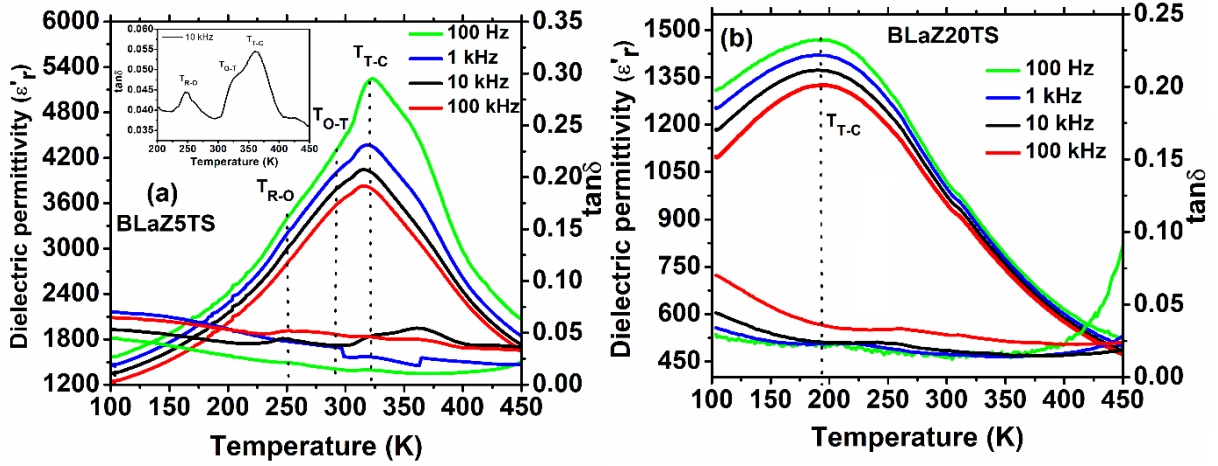

**Figure 4:** Temperature dependence of the dielectric constant and dielectric loss of (a) BLaZ5TS and (b) BLaZ20TS compositions. The inset presents the magnified dielectric losses at 10 kHz showing the three-dielectric transition of the BLaZ5TS composition.

3.4. Ferroelectric and energy storage investigations

To study the ferroelectric behavior of the BLaZ5TS material, P–E hysteresis loops at 20 Hz were recorded at different temperatures (123 K – 423 K). Figure 5 (a, b) shows the polarization versus the electric field of BLaZ5TS ceramic. At low temperatures (Figure 5 (a)), we observe typical ferroelectric behaviors of the P-E loops which are characteristic of the rhombohedral structure (T < 250 K), as observed from the dielectric investigations. The maximum polarization ($P_m$) of the P-E loops (T < 250 K) is seen to increase with temperature. This behavior has been reported widely in ferroelectric materials [57,58], where the hysteresis loops are measured near a FE-FE phase transition. Remind that at room temperature, the system evolves to a tetragonal structure (*P4mm*), where the orthorhombic to tetragonal phase transition occurs around 292°C. The thermal evolution of the P-E loops reveals a narrowing of the loop, with a decrease in both the remanent polarization ($P_r$) and the coercive field ($E_c$) as the temperature increases. At room temperature, the narrowing is even more pronounced due to the structural change. The narrowing of the P-E loops while keeping a relatively high maximum polarization is recommended for energy storage applications.

To investigate the energy storage properties of the BLaZ5TS system, we presented in figure 5 (c) the values of the polarization ($\Delta P = P_m - P_r$), and the coercive field as a function of temperature. Notice that the energy storage main parameters are the energy storage density ($W_{rec}$), and the energy efficiency ($\eta$) and can be written as follow:

$$W_{rec} = \int_{P_r}^{P_m} EdP \qquad 3$$

$$\eta = \frac{W_{rec}}{W_{rec}+W_{loss}} \times 100 \qquad 4$$

Where $W_{loss}$ are the losses dissipated as heat from the system and correspond to the enclosed area of the hysteresis loops. From equation 3, we can conclude that a higher ΔP value, in addition to high applied electric field, are needed for enhanced $W_{rec}$ value. Remind that the ability of the sample to withstand high applied electric fields depends on its microstructure. In fact, the very low grain size obtained for this sample (<1μm) are suitable for the application of a high electric field.

Equation 4 states that for high-energy efficiency, the energy losses should be negligible compared to the energy density values. Thence slimmer is the loop (low $E_c$ and $P_r$ values), higher will be the energy efficiency for a given system. Therefore, from the evolution of ΔP and $E_c$ values, one can get an idea about the energy storage properties of the system.

As observed, ΔP value is increasing with temperature to reach a maximum value around room temperature, and thereafter, decreases monotonously with increasing the temperature. Besides, the coercive field is seen to have a decreasing trend, from low temperature to 350 K, after which the values are seen to stabilize. Thence the highest energy storage density is expected at room temperature with a corresponding relatively high-energy efficiency.

Figure 5 (d) presents the energy storage density and energy efficiency values as a function of temperature. The maximum $W_{rec}$ value of 65 mJ/cm³ is obtained at 12 kV/cm and at room temperature with a corresponding η value of 60%. Owing to the narrowing behavior of the P-E loops with increasing the temperature, the $W_{loss}$ is considerably decreasing and results in higher energy efficiency for the highest temperature (η = 80% for T > 398 K). Whereas, the decrease of $P_m$ behavior with increasing temperature, results in lower $W_{rec}$ values at high temperatures ($W_{rec}$ = 20 mJ/cm³). The obtained results are comparable with results obtained on BCZT systems, with values of $W_{rec}$ = 65 mJ/cm³ and a corresponding efficiency of η = 60% at 300 K. Table 4, gathers the energy storage parameters of some other systems [59–62]. The temperature instability of the energy storage parameters was evaluated using the equation 5 [63]:

$$\eta_{W_{rec}} = \frac{W_{rec_m} - W_{rec}}{W_{rec_m}} \times 100 \qquad 5$$

It should be noted that a relatively low instability ($\eta_{W_{rec}}$ < 20%) is obtained around room temperature between 250 K and 350 K, which is highly recommended for applications.

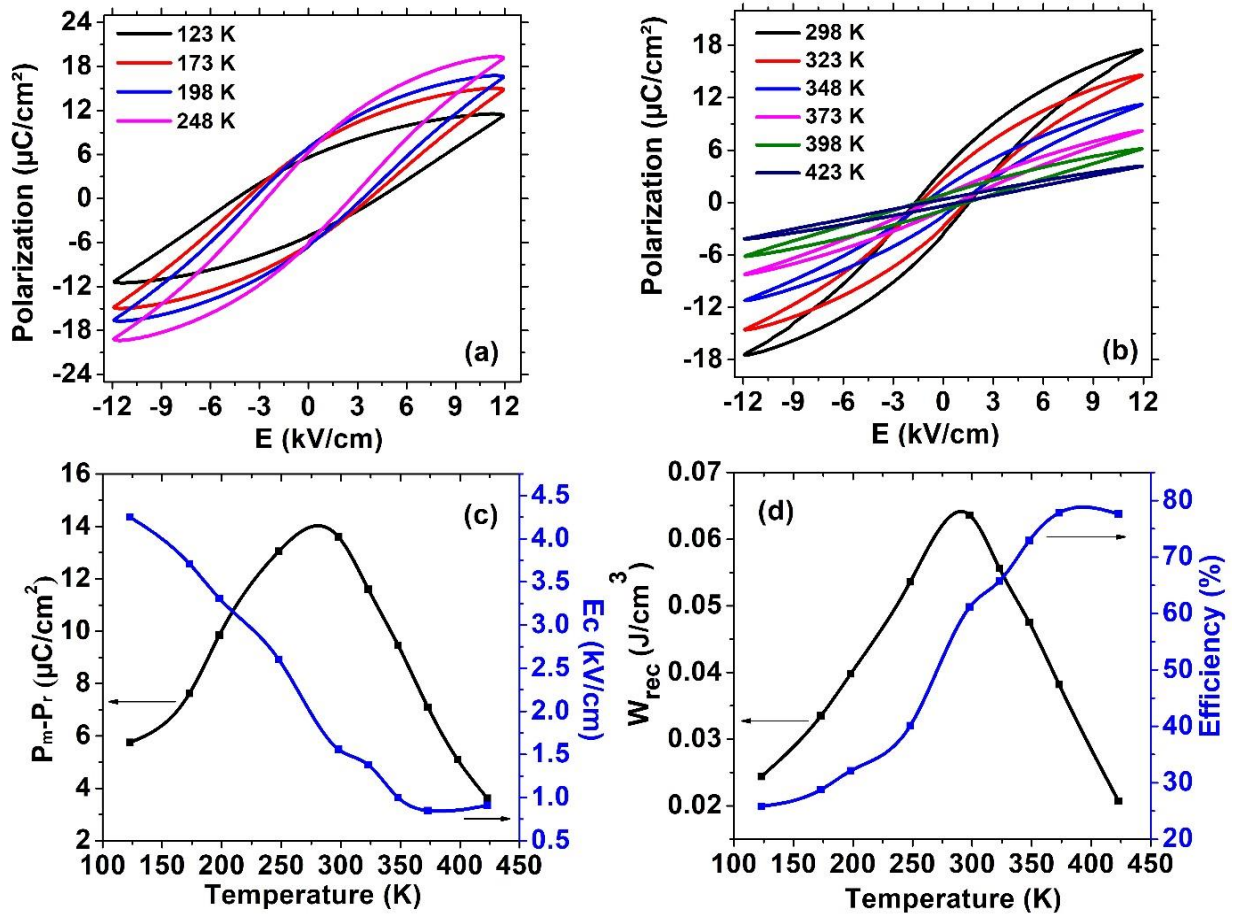

**Figure 5:** (a, b) The hysteresis loops of BLaTZ5S ceramic at different temperatures (123 K - 423 K). (c) The $P_m$-$P_r$ and coercive field as a function of temperature. (d) The energy storage density and energy efficiency versus the temperature.

**Table 4:** Comparison of the energy storage parameters in our work with some literature data.

| Composition | E (kV/cm) | $W_{rec}$ (mJ/cm$^3$) | η (%) | Refs |
|---|---|---|---|---|
| $Ba_{0.85}Ca_{0.15}Zr_{0.10}Ti_{0.90}O_3$ | 6.5 | 14 | 80 | [59] |
| $Ba_{0.94}Ca_{0.06}Zr_{0.16}Ti_{0.84}O_3$ | 120 | 373 | 72 | [60] |
| $Na_{0.5}Bi_{0.48}Gd_{0.02}TiO_3$ | 90 | 850 | 65 | [61] |
| $[(BaZr_{0.2}Ti_{0.80})O_3]_{0.9}[(Ba_{0.70}Ca_{0.30})TiO_3]_{0.1}$ | 80 | 310 | 35 | [27] |
| $Ba_{0.975}La_{0.017}(Zr_{0.05}Ti_{0.90})Sn_{0.05}O_3$ | 12 | 65 | 60 | This work |

3.5. Electrocaloric investigations

The evaluation of the electrocaloric properties of the BLaZ5TS system was done through the calculation of the electrocaloric change of temperature, ΔT. The former can be calculated using the indirect method based on the Maxwell equation:

$$\Delta T = -\frac{1}{\rho C_p} \int_{E_1}^{E_2} T \left(\frac{\partial P}{\partial T}\right) dE \qquad 6$$

Where $\rho$ and $C_p$ are the density (4.2 g/cm$^3$) and the specific heat capacity (430 J/K Kg) of the BLaZ5TS system, respectively. $E_1$ and $E_2$ correspond to the starting and final applied electric fields, and P represents the polarization of the sample. Following the method reported by Mischenko et al. [64], we succeeded in the evaluation of $(\partial P/\partial T)_E$ data. The temperature evolution of ΔT values was plotted in Figure 6 (a), for different applied electric fields (4 – 12 kV/cm). As it can be seen, ΔT shows negative values for temperatures below 250 K. This last behavior can be explained by the increase of the polarization value around the FE–FE phase transition (bump at 250 K in Figure 6 (b)), giving rise to a positive $(\partial P/\partial T)_E$ coefficient which results in a negative ΔT value. Besides, two peaks can be seen in ΔT curves, a first one around 250 K, assimilated to the rhombohedral to the orthorhombic phase transition, and a second peak maximum around 350 K corresponding to the FE–PE phase transition. Naturally, higher is the applied electric field, higher will get the ΔT values. The maximum electrocaloric temperature change value was obtained at 335 K with ΔT = 0.243 K at 12 kV/cm. Remind that the three phase transitions (R-O, O-T, and T-C) are gathered around room temperature for this composition. This special feature is very attractive for ECE applications, since the ΔT peak maxima resulting from the former phase transitions are observed to form a plateau like giving rise to a temperature stability over a large temperature range (250K – 375K).

In order to evaluate the electrocaloric effect of the system without taking into account the applied electric field, the electrocaloric responsivity (ΔT/ΔE) is calculated. Relatively high value of ΔT/ΔE = 0.2 10$^{-6}$ K m/V is obtained, which is comparable with other BCZT based systems (ΔT/ΔE = 0.164 10$^{-6}$ K m/V at 363 K) [18–23,59], and can be even higher to some NBT based systems (ΔT/ΔE = 0.08 10$^{-6}$ K m/V) [61]. Thence, the BLaZ5TS system can be considered as a potential environmentally friendly candidate for solid-state cooling devices. The obtained parameters in this work are gathered in Table 5, along with a comparison with some recent literature data in equivalent systems.

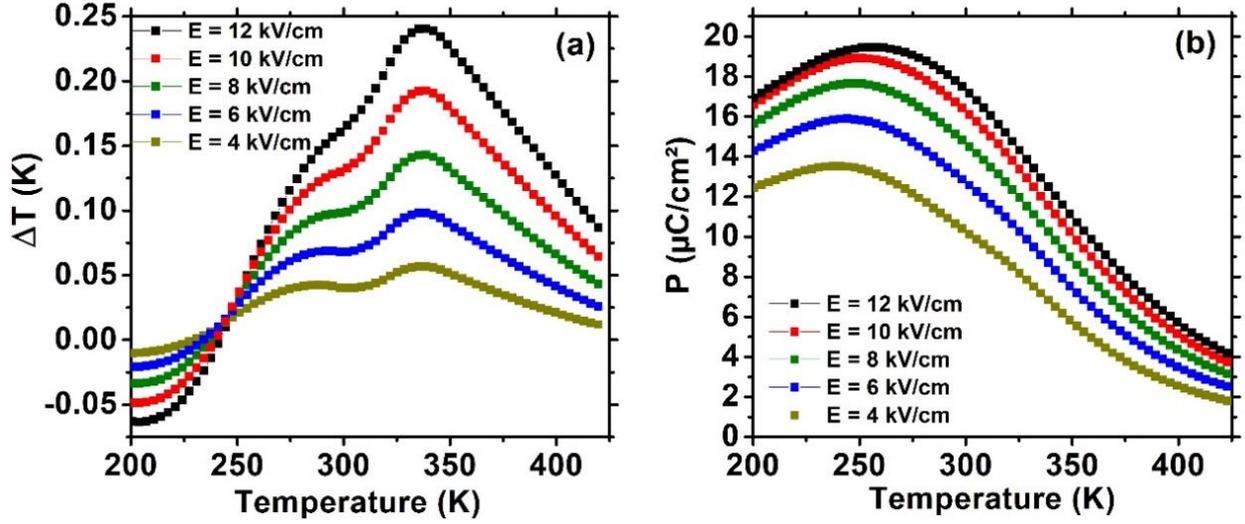

**Figure 6:** (a) The Electrocaloric temperature change ΔT as a function of temperature, (b) the polarization versus the temperature, at different measured electric fields.

**Table 5:** Comparison of the ECE parameters in our work with some literature data.

| Composition | T (K) | ΔT (K) | ΔT/ ΔE × 10$^{-6}$ (K m/V) | Refs |
|---|---|---|---|---|
| $BaTi_{0.95}Sn_{0.05}O_3$ | 358 | 0.079 | 0.19 | [65] |
| $Ba_{0.65}Sr_{0.35}Ti_{0.997}Mn_{0.003}O_3$ | 293 | 3.1 | 0.23 | [66] |
| $Ba_{0.8}Ca_{0.2}Zr_{0.04}Ti_{0.96}O_3$ | 386 | 0.27 | 0.34 | [18] |
| $Ba_{0.98}Ca_{0.02}Zr_{0.085}Ti_{0.915}O_3$ | 358 | 0.6 | 0.15 | [67] |
| $Ba_{0.85}Ca_{0.15}Zr_{0.10}Ti_{0.90}O_3$ | 363 | 0.109 | 0.164 | [59] |
| $Na_{0.5}Bi_{0.48}Gd_{0.02}TiO_3$ | 380 | 0.7 | 0.08 | [61] |
| $Ba_{0.975}La_{0.017}(Zr_{0.05}Ti_{0.90})Sn_{0.05}O_3$ | 338 | 0.24 | 0.20 | This work |

## 4. Conclusions

In summary, BLaZ5TS and BLaZ20TS ceramics have been synthesized by conventional solid-state reaction method and characterized by SEM, XRD, Raman, dielectric, and ferroelectric studies.

SEM images showed samples with high density and low porosity. Compared to BZT, the grain growth seems to be inhibited, and then the grain size is reduced when barium is replaced by lanthanum. The structural refinements underlined a tetragonal symmetry with normal ferroelectric behavior for BLaZ5TS and a cubic symmetry with a diffuse ferroelectric character for BLaZ20TS composition. Raman spectroscopy confirmed the formed phases. The maximum of the permittivity appears at relatively high temperatures for the BLaZ5TS compositions and very low temperatures for the BLaZ20TS compositions.

The energy storage properties have been investigated via the polarization measurements as a function of the temperature for BLaZ5TS. The obtained values are comparable with results obtained on BCZT system, with values of $W_{rec} = 65$ mJ/cm$^3$ and a corresponding efficiency of $\eta = 60\%$ at 300 K. Interestingly, the electrocaloric responsivity was about $0.2 \times 10^{-6}$ K m/V close to the room temperature, which makes this composition a potential candidate for electrocaloric applications.

**Acknowledgments**

The authors gratefully acknowledge the financial support of the "Program Hubert Curien" (PHC-Tassili No. 18MDU107/39978VH); the Haute France Region/ FEDER (project MASENE), and the European Union's Horizon 2020 research and innovation program ENGIMA.
The SEM investigations were conducted using Electron Microscopy facility at University of Picardie Jules Verne (UPJV), Amiens, France.